\documentclass{llncs}

\usepackage{amsmath}
\usepackage{graphicx}
\usepackage{color}
\usepackage{url}
\usepackage{cite}
\usepackage{booktabs} 
\usepackage{subfigure}

\usepackage{amsmath}

\DeclareGraphicsExtensions{.pdf,.png,.jpg}

\usepackage[english]{babel}

\begin{document}
\frontmatter          
\pagestyle{empty}  

\title{Mobile Localization Techniques Oriented to Tangible Web}
 \author{Osvaldo Gervasi\inst{1}$^{ORCID: 0000-0003-4327-520X}$ \and Martina Fortunelli\inst{2} \and Riccardo Magni\inst{2} \and Damiano Perri\inst{1}$^{ORCID: 0000-0001-6815-6659}$ \and Marco Simonetti\inst{3}
 }
%

%
\institute{University of Perugia, Dept. of Mathematics and Computer Science,
Perugia, Italy\\
\and
Pragma Engineering SrL, Perugia, Italy
\and
Liceo Scientifico e Artistico ``G. Marconi'', Foligno, Italy
}
\titlerunning{Mobile Localization Techniques Oriented to Tangible Web} 
\authorrunning{O. Gervasi, M. Fortunelli, R. Magni, D. Perri and M. Simonetti}

\maketitle
\begin{abstract}
We implemented a system able to locate people indoor, with the purpose of providing  assistive services. Such approach is particularly important for the Art, for providing information on exhibitions, art galleries and museums, and to allow the access to the cultural heritage patrimony to people with disabilities.

The system may provide also very important information and input to elderly people, helping them to perceive more deeply the reality and the beauty of art.

The system is based on Beacons, very small and low power consumption devices, and Human Body Communication protocols.
The Beacons, Bluetooth Low Energy devices, allow to obtain a position information related to predetermined reference points, and
through proximity algorithms, locate a person or an object of interest. 

The position obtained has an error that depends from the interferences present in the area. The union of Beacons with Human Body Communication, a recent wireless technology that exploits the human body as a transmission channel, makes it possible to increase the accuracy of localization.

The basic idea is to exploit the localization derived from Beacons to start a search for an electrical signal transmitted by the human body and to distinguish the position according to the information contained in the signal.
The signal is transmitted by capacitance to the human body and revealed by a special resonant circuit (antenna) adapted to the microphone input of the mobile device.

\end{abstract}

\section{Introduction}

The proliferation of wireless devices in the last years has resulted in a wide range of services including indoor localization\cite{indoorSurvey}. 
Indoor localization is the process of obtaining a device or user location in an indoor setting or environment. 
Indoor device localization has been extensively investigated over the last few
decades, mainly in industrial settings and for wireless sensor networks and robotics. However, it is only less than a decade
ago since the wide-scale proliferation of smart phones and wearable devices with wireless communication capabilities
have made the localization and tracking of such devices synonym to the localization and tracking of the corresponding
users and enabled a wide range of related applications and services. 

User and device localization have wide-scale applications in health sector, industry, disaster management\cite{disasterManagement},
building management, surveillance, and other innovative areas as Internet of Things (IoT)\cite{localisationIoT}, smart
cities\cite{localisationSmartCity} and smart environments.

In most devices there is a GPS receiver that allows to easily locate, by satellite signal, a terminal in an open space; unfortunately, for indoor clients most services become unavailable due to the lack of (or a weak) communication signal. It is important to emphasize that, depending on the information we want to obtain, there are two methods for identifying the position of a client: localization and proximity.

Localization is a method used mainly in navigation in open space (outdoor) using the GPS signal: this technology allows obtaining precise information on the user's position. The satellite signal cannot be transmitted inside buildings, and is therefore unusable indoor: this has led to the development of alternative technologies, such as radio, infrared, ultrasound or magnetic fields.

The results obtained with these techniques does not provide an absolute position data, but relative position information is provided, that has then to be interpreted to provide a reliable position. Therefore, these technologies are identified with the term \textit{proximity techniques}. In this context it is necessary to calculate the distance from the various points of interest that will lead to determining, after the execution of certain algorithms, the position sought\cite{fmRadioFingerprint}.

The state of the art on the subject of indoor positioning and tracking is such that there is not a single technology that appears to dominate, 
but several technologies are adopted, each characterized by advantages and disadvantages. The main technologies for indoor localization are described in the table \ref{tab:tech}.

\begin{table}
 {\small
\centering
\caption{Overview of indoor technologies\label{tab:tech}}
\begin{tabular}{||p{0.2\textwidth}|p{0.3\textwidth}|p{0.4\textwidth}||}\hline\hline
	{\footnotesize \textbf{Technology}} & \bf {\footnotesize \textbf{Advantages}} & \textbf{ \footnotesize Disadvantages} \\\hline\hline
	{\footnotesize \textbf{Infrared}} &  {\footnotesize low cost} & {\footnotesize Short range and need to maintain visual contact } \\\hline
	{\footnotesize \textbf{Bluetooth}} & {\footnotesize Infrastructure availability} & {\footnotesize Channel saturation in the case of a large number of users per unit area} \\\hline
	{\footnotesize \textbf{Ultrasounds}} & {\footnotesize Limited battery consumption and poor maintenance} & {\footnotesize Reflection of sound waves in the presence of obstacles } \\\hline
	{\footnotesize \textbf{RFID active}} & {\footnotesize Reading from great distances} & {\footnotesize Fairly high costs, need for power supply, large size} \\\hline
	{\footnotesize \textbf{RFID passive}} & {\footnotesize Infinite average life, reduced cost, small dimensions and resistant to external shocks and stresses} & {\footnotesize Radius and reduced reading range} \\\hline
	{\footnotesize \textbf{Dead-reckoning}} & {\footnotesize Low cost} & {\footnotesize High energy consumption and low range} \\\hline
	{\footnotesize \textbf{Wifi}} & {\footnotesize Infrastructure availability} & {\footnotesize Low accuracy, channel saturation, poor information security} \\\hline
	{\footnotesize \textbf{Ultra-wide band}} & {\footnotesize Accuracy of the order of 1m, robust to multi-path phenomena, does not require visual contact} & {\footnotesize Very high costs} \\\hline
	{\footnotesize \textbf{Zigbee}} & {\footnotesize Low cost, long battery life and accuracy of the order of 1m} & {\footnotesize Limited penetration of the walls } \\\hline
	{\footnotesize \textbf{Bluetooth Low Energy}}& {\footnotesize Low energy consumption, reduced costs, large-scale distributed network} & {\footnotesize  unidirectional } \\\hline
\end{tabular}		
}
\end{table}

Various indoor navigation technologies have been tested in order to identify the best one for locating a person indoor, inside a museum or a room, and then being able to trace her/his movements. We selected the Bluetooth Low Energy (BLE) technology, a new protocol designed for reducing the battery consumption and for optimizing the communications among Internet of Things (IoT) devices.

We use Beacons for localizing and then tracing people in indoor areas, transmitting the data to a mobile device, using BLE and the Human Body Communication protocols.
The data transmitted by a BLE Beacon is contained in packets formatted according to the Bluetooth Core Specification.

Applications interact with Beacons in two different ways:
\begin{list}{\ \ \ }{}
 \item \textit{Monitoring}: action activated when entering and leaving the region of Beacon monitoring, works whether the application is running, paused or stopped;
 \item \textit{Ranging}: action activated based on the proximity of the Beacon, it works only when the application is running.
\end{list}

Therefore the monitoring allows to identify the Beacon regions, while the ranging is used to interact with a specific Beacon. 
The standardized Beacon protocols are IBeacon and Eddystone; in our work we adopted the Eddystone protocol, developed by Google Inc. and released under the open source Apache License 2.0.\cite{eddystone} 

The implemented system is relevant to provide assistive services to people, in particular for elderly people, who can take advantage of the system in case they feel lost or disoriented. The system may be also used for general purpose services, in particular to provide specific informations on artifacts in a museum or in an art gallery, art, history and tourism information nearby relevant monuments or popular meeting points in a city or a suburb.

The present work is addressing also the theme of so called \textit{tangible web}\cite{tanbits,digitan} interactions, which are completing the sensory engagement with audio and video of multimedia interfaces. 
The idea to associate a value of haptic exploration in retrieving information (i.e. from web sources) humanises the learning experience offering a contextual feedback which is related to forms and textures sensed by touch. 
The chance to detect such interaction, identifying the specific user's device, allows the access to specific information which is related to the experience that the user is making in a certain moment, adding contextual or related information or feelings (i.e. sounds and music). 

The basis of such developments could be recognised in both afferent and efferent stimulations: in terms of afferent stimula, the current work is making use of recent applications of HBC - Human Body Communication solutions - which are proposing a "natural" paradigm in the relationship between the artifact and the device weared by the user.

The development of such technology will offer the chance to produce emotionally enriched explorations of artifacts also for those who are deprived of vision (low vision-blind persons), as mentioned in the reported use case.


\section{The Human Body Communication}

Human Body Communication (HBC) is a recent wireless technology which is part of the Body Area Network (BAN), able to interconnect wearable devices at distances lower than 1 meter. The human body becomes the communication medium defined as "Body-wire channel" that can propagate frequencies in the range from 10KHz to 100 MHz. 

The human body is made up of ions and can suffer harmful effects depending on the frequency, the intensity of the current, the path followed by the current, and the duration of the interaction\cite{eplasty}]. Generally high frequencies are less dangerous because they are accompanied by a skin effect, in fact the possibility of fibrillation decreases, as the current passes outward without affecting the heart, although at the same time there is a reduction of the impedance on the human body, which determines a current increase at the same voltage.

There are directives, issued by the ICNIRP, the International Commission on Non-Ionizing Radiation Protection, and IEEE, Standard for Safety Levels,  that limit the time of exposure to electric, magnetic and electromagnetic fields. It has been proven that the pain threshold varies between 100kHz and 1MHz. Below 100kHz one can feel a slight stimulation of muscles and nerves, from 100kHz to 10MHz one feels a sensation of heat, while for over 10MHz  one is beyond the percentage of energy that the human body is able to absorb when it is exposed to the action of a radio-frequency electromagnetic field (RF).

In literature, there are two approaches of HBC: galvanic coupling and capacitive coupling. Both methods use body as transmission medium, but they differ in the coupling between signal and human body. In the present work we have analyzed only the capacitive coupling.

\subsection{Capacitive coupling}

In the 1990s, Thomas G. Zimmerman\cite{zimmerman} explained that electrical signals with a frequency lower than 100kHz interacts with the human body. When a very high frequency comes in touch with the body, the electrical energy does not remain confined but propagates around it as if to simulate an antenna.

\begin{figure}[htbp]
	\centering%
	\includegraphics[width=0.9\textwidth]{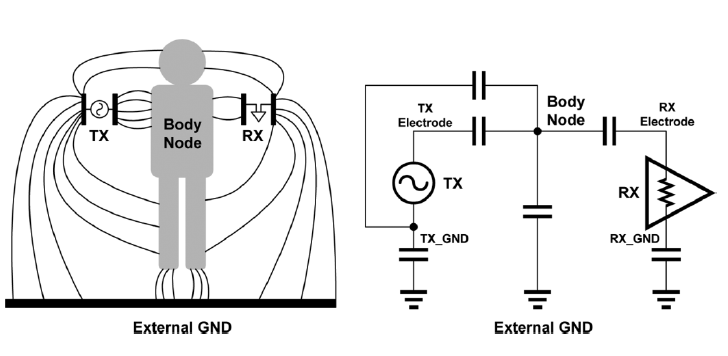}
	\caption{ Model of a Circuit based on Human Body Communication.\label{fig:model}}
\end{figure}

Figure \ref{fig:model} shows the model of a conventional circuit and the near-field coupling mechanism around the human body is described. 
The signal is transmitted between the body channel transceivers by making a current loop, which is composed of the transmitter electrode, the body channel, the receiver electrode, and the capacitive return path through the external ground. As the communication frequency increases for a high data rate, the coupling capacitances of the return path have less effect and the body impedance cannot be ignored. As the transmission length of the body channel increases, both the resistance of the body and the coupling capacitance to the external ground increase. These elements cause signal loss at the receiver, and its amount depends on the channel length.

The Zimmerman's theses is the base reference for further studies, that are different for the amplitude of the coupling, the frequency range, the signal modulation method and the speed of data transmission.

Capacitive coupling has several weaknesses:
\begin{itemize}
	\item The return path of the signal must be managed. The ground conduction is the base for the transmission.
	\item The transmission channel of the dominant signal is on the surface of the arm, because the signal is distributed
	like a wave.
	\item The same contacts can emit dispersal fields.
	\item Higher the carrier frequency is, more the transmission by irradiation through the air becomes relevant.		
\end{itemize}

In a recent study Namjun Cho and co-workers\cite{cho} provided two empirical formulas that express the characteristic of the communication channel and of the minimal area of contact as a function of several parameters.

\section{Signal analysis on Mobile environments}

Signal analysis is a process of different check that where apply in many different environments including the Mobile one. Signal Analysis allows to get only the significant part of the signal that represent the data using a device called filter.

A filter removes some unwanted components from a signal, this means removing some frequencies to suppress interfering signals and reduce background noise. There are many different types of filters, classified according to their properties: passive or active, analog or digital, discrete-time or continuous-time etc.. 
In order to apply a digital filter in Mobile environments it is necessary to obtain a digital signal.
The original analog signal has to be sampled, reading the signal at regular time intervals, quantize the signal, and encode it  with discrete values. The types of digital filters analyzed are: Finite Impulse Response (FIR) and Infinite Impulse Response(IIR). We selected the FIR filter, whose impulse response is of finite duration because it settles to zero in finite time. 

The filtered signal is transformed with a digital technique, in a impulse train of 0, when no signal is resent, and 1, when the signal is present, that represents the data after the On-Off keying(OOK). 
On-Off keying is the simplest form of amplitude-shift keying modulation, which variations is in amplitude of a carrier wave. Modulation is the process of varying one or more properties, as amplitude, frequency, phase etc.. , of a periodic waveform, called the carrier signal, with a modulating signal that contains the information that have to be transmitted.

The last part of signal analysis is the check of correctness of the transmitted data. Error revelation is an important technique to auto detect errors between information's data without correct them.         
The cyclic redundancy check (CRC) is an error-detecting code with checksum where the reminder of a polynomial division of the contents data follows some data blocks. 

\begin{figure}[htbp]
	\centering%
	\includegraphics[width=0.9\textwidth]{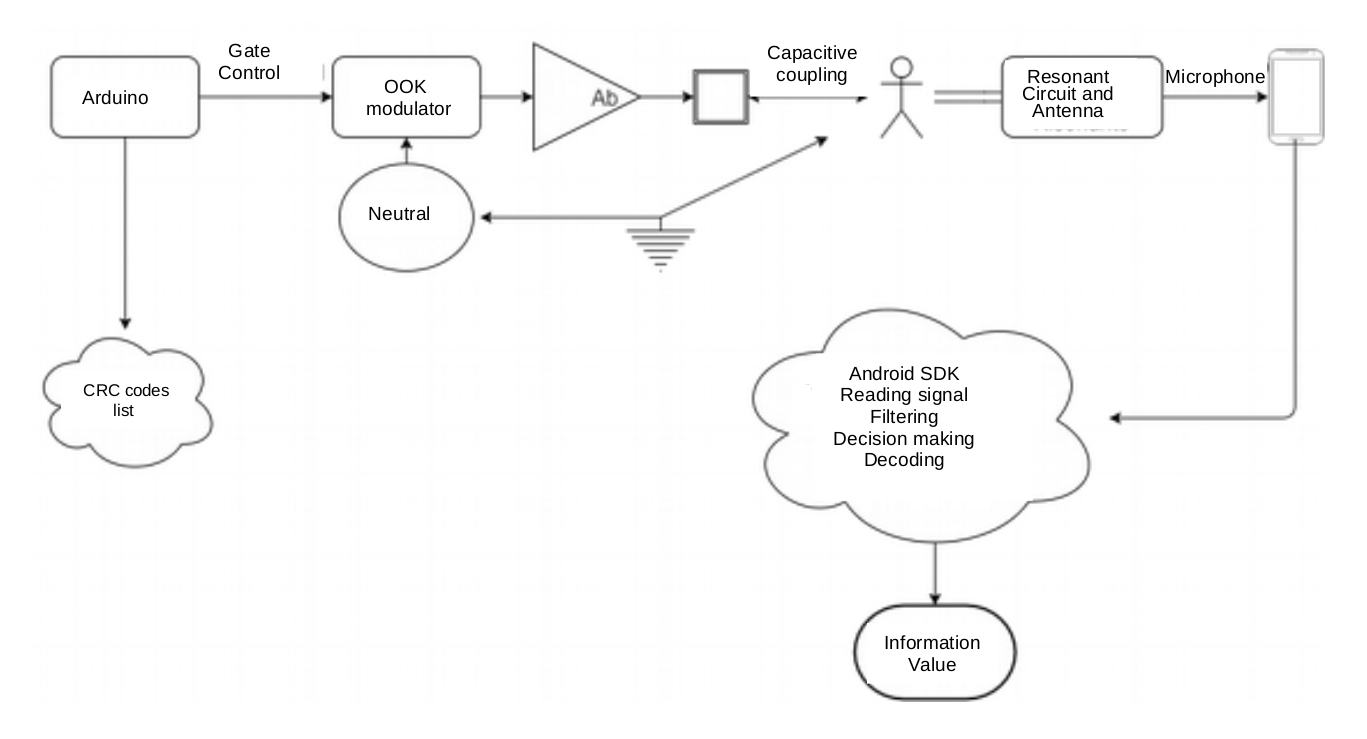}
	\caption{The setup of the experimental apparatus.\label{fig:setup}}
\end{figure}

The described setup is summarized in Figure \ref{fig:setup}.
The components that appear in the Figure are described in Table\ref{tab:setup}.

\begin{table}
  \begin{center}
    {\small
		\centering
		\caption{Description of the Setup elements\label{tab:setup}} 
	\begin{tabular}{||p{0.2\textwidth}|p{0.6\textwidth}||}\hline\hline
		{\footnotesize \textbf{Element }}&{\footnotesize \textbf{Description}}\\\hline\hline
		{\footnotesize \textbf{Arduino}} &{\footnotesize Board Arduino Uno with microcontroller ATmega328, velocity 16MHz, flash memory 32KB, SRAM 2KB and EEPROM memory 1KB. The program generates on the exit pin a bit sequence at 100 bps, modulated to 1KHz.}\\\hline
		{\footnotesize \textbf{Modulator OOK}} & {\footnotesize Wave shapes generator HM130 from 0 to 10 MHz. The square wave generated at 155kHz with  50$\Omega$ input and exit gate} \\\hline
		{\footnotesize \textbf{Ab}} & {\footnotesize HF (High Frequency) frequency amplifier with resonant circuit on exit}\\\hline	
		{\footnotesize \textbf{Neutral}} & {\footnotesize Ground reference}\\\hline
		{\footnotesize \textbf{Resonant Circuit Antenna}} & {\footnotesize  Weared unit connected to the microphone of the mobile phone (dimensions: 50x30x4 mm)}\\\hline
		{\footnotesize \textbf{Mobile Device}} & {\footnotesize  Smartphone Android 4.4 (KitKat)}\\\hline	
	\end{tabular}		
   }     
  \end{center}
\end{table}

Philip Koopman proposed the guidelines for identifying the best polynomial function related to 8 bits payload. In particular he compared 3 different polynomial: DARC-8, CRC-8 and C2. C-2 show the best performance with a frame length of 119 and a payload greater than 10 bits, while DARC-8 showed the best performance with 8 bits payload. So the polynomial function selected was  DARC-8.

\section{The electronic apparatus}

The electronic apparatus developed for data communication implements the interfaces for inductive human body transmission of signals. The scheme provides the emission of low-freq modulation (which is modulating the base band signal) generated under the surface of the sensorized object and then conveyed by the user skin during the contact toward a receiver antenna which is resonating at the same frequency. For sake of simplicity, the modulation frequency has been chosen at 125KHz, allowing the utilization of coils and related components already available for wireless charge systems. Next figure shows the overall block scheme.

\begin{figure}[htbp]
	\centering%
	\includegraphics[width=0.9\textwidth]{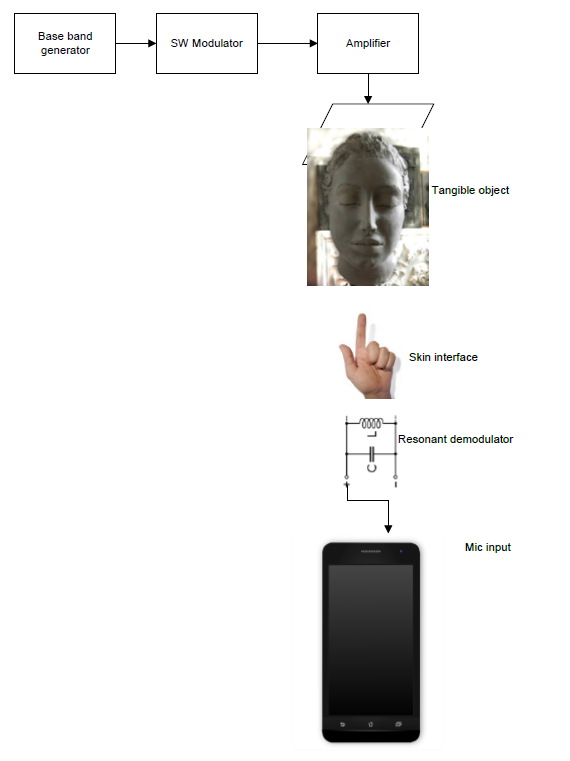}
	\caption{Electronic apparatus \label{fig:risultati}}
\end{figure}

In this conceptual scheme the main blocks are:
\begin{list}{- }{}
\item Baseband generator is constituted by the coded string that is periodically sent to the apparatus. i.e. tangible object unique identifier
\item SW modulator is working with the modulation frequency to realize OOK (On-Off Keying) modulation
\item The amplifier is energizing the plate under the plastic reproduction (i.e. a 3D printed scale copy of the tangible item) 
\item The skin interface conveys the energy coming from the capacitive coupling toward a wrist-placed detection system
\item The resonant demodulator is trimmed around the central modulation frequency producing a demodulated input at low level (2-5 mV)
\item The microphone input of the mobile device is detecting the modulation allowing a the receiver process to recognize data string
\end{list}

\section{The implemented libraries in Android}

The present part is focused on the libraries implemented in Android. The code implemented can be divided in four different parts in which we consider the signal acquisition, the signal filtering, the decisioning and finally the encoding.

The signal is acquired in realtime through the microphone in PCM (Pulse Code Modulation) for \textit{n} seconds. After the recording the signal is filtered with a FIR filter with 20 TAP, that was compared to the one obtain with LabVIEW library. Results of many tests underline that the use of filter is not necessary because rumor doesn't hold the signal's presence or absence.

The third part check if in an instant \textit{t} the signal is present or not. 
After bringing all the samples positive, it is applied an average function based on a window that take \textit{k} samples with some overlays in order to obtain a greater precision. The window dimension can be determined only through experiments. The results are then "decimated" with a technique in which the samples occurring in the same millisecond are reduced to one. The "decimating" algorithm reads tree different values, if one of these is \textquoteleft 1\textquoteright, it considers that there is signal. Also the distance between the tree different samples is based on experimental tests. The aim of "decimating" algorithm is to read the central part of a millisecond, in order to have an high probability of take the correct measure.

Finally the correctness of the result is analysed and  the reliability of the transmission is verified, using the Cyclic Redundancy Check (CRC), detecting the frames trasmitted with errors.

\section{The use case: Art for everyone}

Many use cases of this project can be identified in everyday life, in particular the implemented system may help elderly people and people with disabilities, such as blind or hypovision people, who are for example excluded from the visual art.
The only way for such people to access visual art is through the tactile and auditory sensations and taking into account this type of disabilities we took the cue to carry out this project.

Let us take an archaeological museum as an example, the visitor with disabilities may be equipped with the headset and sensors necessary to capture the signal transmitted through the human body connection to the microphone jack.  The application may be installed in the personal mobile device of the visitor so that only the headset and the antenna should be provided to her/him.

The description of the artifacts may start as soon as the visitor is intercepted by a beacon and the related information transmitted. Once the artifact is identified, the application starts an audio that describes it, and if more sensors have been installed, the audio may describe the area the visitor is touching.

The use of the application can be extended to painting, photography, architecture, and all visual arts.


\section{Conclusions and future work}

The present work describes a system able to localize  people indoor and is particularly important for helping elderly people in case they feel lost or disoriented, and people with disabilities helping them receiving information. 

This work is another pillar on the way we started several years ago, helping people with disabilities to recover in part the limitations originated by their health status\cite{gervasi2010,gervasi2008,gervasiICCSA2016,gervasiICCSA2017,gervasiICCSA2015,gervasiICCSA2014,gervasi2019}.

The system may be improved optimizing the number of beacons required to cover indoor areas and to improve its precision adopting triangulation methods.



\bibliographystyle {plain}
\bibliography {main}

\end{document}